\documentclass[twocolumn,prd,graphicx,showpacs]{revtex4}

\usepackage{graphicx}
\usepackage{dcolumn}
\usepackage{bm}

\newcommand{\be}{\begin{equation}}
\newcommand{\ee}{\end{equation}}
\newcommand{\ba}{\begin{eqnarray}}
\newcommand{\ea}{\end{eqnarray}}

\newcommand{\lsim}   {\mathrel{\mathop{\kern 0pt \rlap
  {\raise.2ex\hbox{$<$}}}
  \lower.9ex\hbox{\kern-.190em $\sim$}}}
\newcommand{\gsim}   {\mathrel{\mathop{\kern 0pt \rlap
  {\raise.2ex\hbox{$>$}}}
  \lower.9ex\hbox{\kern-.190em $\sim$}}}

\begin{document}
\title{New hadrons as ultra-high energy cosmic rays}
\author{Michael ~Kachelrie{\ss}$^a$, Dmitry~V.~Semikoz$^{a,b}$ and
Maria A. T\'ortola$^c$}

\affiliation{
$^a$Max-Planck-Institut f\"ur Physik (Werner-Heisenberg-Institut),
F\"ohringer Ring 6, D--80805 M\"unchen, Germany\\
$^b$ Institute for Nuclear Research of the Academy
of Sciences of Russia, Moscow, 117312, Russia\\
$^c$ Instituto de Fisica Corpuscular - C.S.I.C., Universitat de Valencia,
Edificio Institutos de Paterna, Apt. 22085, E-46071 Valencia, Spain}

\begin{abstract}
Ultra-high energy cosmic ray (UHECR) protons produced by uniformly
distributed astrophysical sources contradict the energy spectrum
measured by {\em both\/} the AGASA and HiRes experiments, assuming the
small scale clustering of UHECR observed by AGASA is caused by
point-like sources.  In that case, the small number of sources leads
to a sharp exponential cutoff at the energy $E<10^{20}$~eV in the
UHECR spectrum. New hadrons with mass 1.5--3~GeV can solve this cutoff
problem. For the first time we discuss the production of such hadrons
in proton collisions with infrared/optical photons in astrophysical
sources. This production mechanism, in contrast to proton-proton
collisions, requires the acceleration of protons only to energies $E
\lsim 10^{21}$~eV. The diffuse gamma-ray and neutrino fluxes in this
model obey all existing experimental limits. We predict large UHE
neutrino fluxes well above the sensitivity of the next generation of
high-energy  neutrino experiments. As an example we study hadrons
containing a light bottom squark. This model can be tested by accelerator
experiments, UHECR observatories and neutrino telescopes.
\end{abstract}

\pacs{
14.80.-j,   
98.70.Sa    
}

\date{February 18, 2003}

\maketitle

\section{Introduction}
UHECRs with energies above $10^{19}$~eV have been observed in all relevant 
experiments, i.e. Volcano Ranch \cite{Volcano}, Haverah
Park~\cite{Haverah}, Fly's Eye~\cite{Fly},  
Yakutsk~\cite{Yakutsk}, AGASA~\cite{agasa} and HiRes~\cite{Hires}. 
Their arrival directions are distributed uniformly over the sky
without significant correlation with the galactic or the super-galactic 
plane. This isotropic distribution is consistent with the simplest
model for UHE primaries, 
in which protons are accelerated in extragalactic, uniformly
distributed astrophysical sources. However, UHE protons with
energies above $E>4 \times 10^{19}$~eV interact with cosmic microwave
background (CMB) photons and lose quickly energy through pion
production within 50 Mpc. As a consequence, a cutoff in the UHECR
spectrum, predicted already in 1966  by Greisen, Zatsepin and Kuzmin 
(GZK)~\cite{gzk}, should show up for uniformly distributed sources 
at $\sim 5\times 10^{19}$~eV. 
This cutoff is not observed by the ground array experiment with the
largest exposure, AGASA, while the first monocular results of the HiRes
fluorescence telescope are in agreement with the GZK cutoff.  
The exposure at the highest energies of all other experiments is too
small to allow for a definite conclusion about the presence or absence of
the GZK cutoff.

Fortunately, the next generation experiment Pierre Auger Observatory
\cite{auger} which is a combination of an array of charged particles
detectors with several fluorescence telescopes is currently under
construction. It will not only be able to resolve possible
systematic differences between the ground array and fluorescence
telescope techniques, but will also increase the statistics of UHECR
data by an order of magnitude. The telescope array project, also based on the
fluorescence technique, may serve as the optical component of the planned
northern Pierre Auger site~\cite{ta}. There are also plans for space
based observatories such as EUSO~\cite{euso} and OWL~\cite{owl} with
even bigger acceptance.  

Assuming that the GZK cutoff will be confirmed
by future experiments does not resolve the UHECR puzzle. Since all
experiments including HiRes see events with energies $E>10^{20}$ eV, their
sources should be located within the distance $R\lsim 50$~Mpc.
Otherwise, the GZK cutoff is extremely sharp and in contradiction even
to the UHECR spectrum  measured by HiRes (cf. Fig.~\ref{F4}). But
there are not many astrophysical sources within this distance from the 
Earth known able to accelerate particles to the highest
energies. Moreover, these sources are not located in the directions of
observed events. Another problem is a statistically
significant ($4.6~\sigma$ for energies above $4\times 10^{19}$~eV)
clustered component in the arrival directions of AGASA
data~\cite{AGASAcluster_data,AY_cluster_data,AGASAcluster_data2,cluster_more}. 
The
sensitivity of the other experiments for clustering at the energies
$E>4 \times 10^{19}$~eV is much smaller, either because of the
smaller exposure at the highest energies (Yakutsk) or because of a
poor two-dimensional angular resolution (HiRes in monocular mode).
At lower energies $10^{19} ~{\rm eV} < E < 4 \times 10^{19}$~eV a
clustered component still exists in the AGASA
data~\cite{AGASAcluster_data2}, but with a reduced significance of
$2.3~\sigma$. The Yakutsk experiment also observes a clustered
component in the energy region $E > 2.3 \times 10^{19}$~eV
with a chance probability  $2\times 10^{-3}$ or $\sim 3 \sigma$ 
using Gaussian statistics~\cite{AY_cluster_data}.

The puzzle of the GZK cutoff can be solved in two different ways. 
The first one supposes 
that the sources of UHECR are located nearby. Then the extragalactic
magnetic field should be strong enough, $B\gsim 0.3\mu$G, 
to deflect UHECR with $E>10^{20}$ eV and magnetic lensing could be
responsible for the clustered component~\cite{sigl_and_co}. A problem
of this solution is the difficulty to construct realistic maps of the
matter and magnetic field distribution in the nearby Universe. 
Simulations done so far reproduced the energy spectrum and the
clustered component assuming  10--100 sources, but without using
realistic locations of the sources. Another difficulty is that
magnetic lensing, although reproducing the clustered component, 
predicts in general a broad angular distribution of this component,
while the data are within the experimental angular resolution.   
This solution is even more problematic if future experiments show the
absence of the GZK cutoff: fixing the luminosity of the local sources
to the UHECR flux above the GZK cutoff then results in a too large
UHECR flux at lower energies, where sources from the whole Hubble
volume contribute.

The second way is to suppose that the clustered component is due to a
neutral particle which is not deflected by (extra-) galactic fields.
In this case one can look for correlations of UHECR arrival directions
with astrophysical objects.  Tinyakov and Tkachev have recently found 
a significant (more than 4~$\sigma$) correlation with BL
Lacs~\cite{corr_bllac}.  The BL Lacs which correlate with the UHECRs
are located at very large (redshift $z\sim 0.1$)  or unknown distances.
If it can be shown with an increased data set of UHECRs that this
correlations holds also at energies $E>(6-10)\times 10^{19}$~eV, then
protons alone can not explain the UHECR data and a new component in
the UHECR spectrum is needed.

The simplest possibility is that this new component is due to
extremely high energy ($E\gtrsim10^{23}$~eV)  photons emitted by
distant sources. They can propagate several hundred Mpc constantly
losing energy and thereby creating secondary photons also inside the GZK 
volume~\cite{photons}. However, this model requires extremely small
extragalactic magnetic fields, $B<10^{-12}$~G, and the minimal possible
radio background.  Besides, one needs to accelerate protons to
$E\gtrsim10^{24}$~eV in order to create such photons.
An acceleration mechanism to these extreme energies is not known.

Another possibility is that the events beyond the GZK cutoff
are related to the $Z$ burst model~\cite{zburst1}. In this model,
UHE neutrinos interact with the relic neutrino background producing
via the $Z$ resonance secondary protons and photons.
The big drawback of this scenario is the need of an enormous flux of
primary neutrinos that cannot be produced by  astrophysical
acceleration sources without overproducing the GeV
photon background~\cite{kkss}. Also in this model,
primary protons have to be accelerated to extremely high energies, 
$E\gtrsim10^{23}$~eV, in order to produce $E=10^{22}$ eV neutrinos.

Conventionally, acceleration mechanisms allow to accelerate protons in
astrophysical sources only up to $E \lsim 10^{21}$~eV. If one considers
this maximal energy as a serious upper limit, both possibilities
discussed above are excluded and some kind of new particle physics
beyond the standard model is required. The most radical option is
violation of Lorentz invariance~\cite{LV}. A more conservative, though
for same tastes still too speculative, possibility are decaying
super-heavy relics from the early Universe~\cite{SDM}. This model
cannot explain the correlation of UHECR arrival directions with BL
Lacs, and could be excluded if these correlations are found also at
energies $E\gsim (6-10)\times 10^{19}$~eV. 
In theories with a fundamental scale of gravity as low as
${\cal O}$(TeV), the interactions of neutrinos with nucleons are
enhanced compared to the standard model by the exchange of
Kaluza-Klein gravitons~\cite{shrock} or the production of black
holes~\cite{BH}. In Refs.~\cite{claim} it was claimed that neutrinos could 
have hadron-like cross sections at UHE and be responsible for the observed
vertical air showers. However, unitarization slows down the growth of
the neutrino-nucleon cross section~\cite{MP,GRW} and moreover, in the
case of exchange of Kaluza-Klein gravitons, the energy transfer to the
target nucleon is too small~\cite{MP}.  Thus the neutrino-nucleon
cross-section can be up to a factor 100 larger than in the standard
model, but neutrinos in both cases still
resemble deeply penetrating particles and cannot imitate air showers
initiated by nucleons.

Another possibility is that new particles are directly produced in
astrophysical sources. A model with an axion-like particle, i.e. a 
scalar which can mix with a photon in the presence of
external magnetic fields, was suggested in  Ref.~\cite{axion}. 
Axion-like particles can be also produced by photons  emitted by 
astrophysical sources  via axion-photon oscillations~\cite{Csaki:2003ef}. 
In supersymmetric (SUSY) theories  
with conservation of $R$ parity, the lightest supersymmetric particle
(LSP) is stable and can be a HE primary. This possibility was for the
first time seriously discussed in connection with Cyg X-3 in the
80's~\cite{X3,BI86}. More recently, the production and
interactions of both the neutralino and the gluino as LSP at UHE were 
examined in Ref.~\cite{Be98}. The authors concluded that only a
light gluino could be produced in reasonable amounts by astrophysical
accelerators. Reference~\cite{neutralino} calculated the neutralino
production in proton-proton collisions and found that
neutralinos produced in astrophysical sources cannot be an important
UHE primary: Since the production cross section of neutralinos is too
small, this model predicts either a negligible flux of UHE
neutralinos or is not consistent with measurements of the diffuse gamma-ray
background~\cite{egret} and with existing limits on the neutrino flux
at ultra-high energies. The latter limit were obtained by the
Fly's Eye~\cite{Fly'sEye_nu}, AGASA~\cite{agasa_nu}, RICE~\cite{rice},
and GLUE~\cite{glue} experiments.

SUSY models with a strongly interacting particle as LSP or  next-to-lightest 
SUSY particle (NLSP) are more interesting for UHECR physics. 
Hadrons containing a gluino were first
suggested by Farrar as UHECR primary~\cite{Fa96,farrar}. Her model, a light
gluino $\tilde g$ together with a light photino such that the photino
could serve as cold dark matter candidate, is excluded~\cite{exp_g1,exp_g2}. 
Motivated by the correlation of UHECR with BL Lacs, the production of
light $\tilde gg$ bound-states in astrophysical accelerators
was suggested in Ref.~\cite{Berezinsky:2001fy}.
However, the light gluino window seems to be now closed by
Ref.~\cite{Janot} also for generic models with a light gluino.

In this paper, we start from a model-independent, purely
phenomenological point of view. Since the observed extensive air
showers (EAS) are consistent with simulated EAS initiated by protons,
any new primary proposed to solve the GZK puzzle has to produce EAS
similar to those of protons. An experimentally still open possibility
are photons as UHECR primaries: 
at 90\% C.L., $\sim 30\%$ of the UHECR above $E>10^{19}$~eV can be
photons~\cite{photon_limit}.  However, the
simplest possibility consistent with air shower observations  
is to require that a new primary is strongly interacting.
The requirements of efficient production in astrophysical accelerators
as well as proton-like EAS in the atmosphere ask for a
light hadron, $\lsim 3$~GeV, while shifting the GZK cutoff to
higher energies results in a lower bound for its mass,
$\gsim 1.5$~GeV~\cite{Berezinsky:2001fy}. From these requirements, we
derive general conditions on the interactions of new UHE primaries.

As specific example, we investigate the case of a bottom squark
containing hadron which we call ``shadron'' from now on. Our
conclusions are however independent of the underlying particle physics
model for the shadron. The required properties of the shadron will be
parameterised as function of its mass, production and
interaction cross-sections. Other suitable realizations of shadrons
could 
be new (meta-) stable hadronic states like H-dibaryons~\cite{H}.  
Another candidate could be connected to the exotic, charged hadronic
state with mass 2.3~GeV discovered recently by BaBar~\cite{BaBar} 
which was suggested by the experiment as a four-quark state. If this
suggestion would be true, then this four-quark state could be related to a
new metastable neutral hadron.  

We find that proton-proton
collisions in astrophysical accelerator cannot produce high enough
fluxes of new primaries without contradicting existing measurements of
photon~\cite{egret} and neutrino fluxes~\cite{Fly'sEye_nu,agasa_nu,rice}.
By contrast, we find for a light shadron with mass  $\lsim 3$~GeV and
the astrophysically more realistic case of UHE proton collisions on
optical/infrared background photons no contradiction with existing
limits. Also, the required   
initial proton energy is not too extreme, $E \lsim 10^{21}$ eV, which is
compatible with existing acceleration mechanisms. The only essential
condition for the  sources is that they should be  
optically thick for protons in order to produce these new hadrons. (This
condition is similar for all models with new particles produced by
protons). Below we will show that at least some of BL Lacs correlated 
with UHECR obey this condition.

One of the important features of the proposed model, and any model in
which the production cross section $\sigma_{p\gamma\to S}$ of a new
particle $S$ is much smaller than the total proton-photon cross
section $\sigma_{p\gamma}$, is the high flux of secondary high-energy
neutrinos. This neutrino flux is connected via the 
relation  $F_{\rm  CR}\sigma_{p\gamma}/\sigma_{p\gamma\to S}$ to   
the maximal contribution of $S$ particles to the cosmic ray flux,
$F_{\rm CR} \approx 1/E_{20}^2$~eV/(cm$^2$ s sr). It can be
detected by future UHECR experiments like the Pierre Auger Observatory
\cite{auger}, the Telescope Array~\cite{ta}, EUSO~\cite{euso} and
OWL~\cite{owl}. Alternatively, such neutrino fluxes can be detected by 
triggering onto the radio pulses from neutrino-induced air
showers~\cite{radhep}. 
Acoustic detection of neutrino induced interactions is also
being considered~\cite{acoustic}. There are plans to construct telescopes
to detect fluorescence/\v{C}erenkov light  from near-horizontal
showers produced in 
mountain targets by neutrinos  at intermediate energies~\cite{fargion,mount}. 
 Moreover, if the sources are optically thick 
for protons, the neutrino flux can be significant both at
high energies and down to energies $10^{16-17}$ eV, 
depending on the pion-production threshold on optical/infrared photons
\cite{nu_sources}.  Therefore, one may observe neutrinos
from the same sources both by future UHECR experiments
and by neutrino telescopes like AMANDA~\cite{amanda},
ICECUBE~\cite{icecube}, GVD~\cite{GVD}, ANTARES~\cite{antares},
NESTOR~\cite{nestor} or NEMO~\cite{NEMO}.

The paper is organized as follows. We start with a discussion of the
spectrum of UHECR protons produced by a small number of extragalactic
astrophysical sources in Sec.~\ref{protons}. Then we consider models
containing light strongly interacting particles, shadrons, and their 
status. In Sec.~\ref{propagation} we discuss the  propagation of shadrons 
through the Universe. Their interactions in the atmosphere were 
investigated in detail before, so we shall just briefly recall the main
characteristics in Sec.~\ref{atmosphere}. Section~\ref{sources} is
devoted to a detailed analysis of shadron production in astrophysical
sources. In Sec.~\ref{parameters} we discuss all astrophysical
constraints which shadrons have to obey to be viable UHECR
primaries. In Sec.~\ref{sec:BLLac}, we discuss the particular case of
BL Lacs as sources of UHECRs. Finally, we summarize our results in
Section \ref{conclusions}.

\section{Protons from uniformly distributed sources}
\label{protons}

The HiRes experiment published recently their data from monocular
observations~\cite{Hires}. They showed that the UHECR flux is
consistent with the GZK cutoff expected for {\em uniformly,
continuously\/} distributed sources. As a result, the simplest model of
UHECR---protons accelerated in uniformly distributed, extragalactic
sources---seems to be a convincing explanation of their data.
The authors of Ref.~\cite{BGG2002} found as fingerprints of
the expected interactions of UHE protons with CMB photons a
dip at $E\sim 1\times 10^{19}$~eV, a bump~\cite{fingerprints}
and the beginning of the
cutoff in the measured spectra of four UHECR experiments. The
agreement of the spectral shape calculated for protons with the
measured spectra is excellent, apart from an excess in the AGASA data
above $E\gsim 8\times 10^{19}$~eV. These findings
point to an AGN origin of UHECR below $E\lsim 10^{20}$~eV and to
protons as primaries. Despite the fact that the AGASA experiment sees
a significant number of events above the GZK cutoff~\cite{agasa}, the
model of proton primaries from extragalactic sources looks
very attractive, because it does not require new physics.

The model of uniformly, {\em continuously\/} distributed sources is
based on the assumption that the number of UHECR sources is so large
that a significant fraction of sources is inside the GZK
volume. However, as it was shown in a number of
works~\cite{Dubovsky2000}, \cite{AGASAcluster_data2}, \cite{FK2000}, 
the small scale clustering of UHECR observed by AGASA allows to
estimate the number of UHECR sources assuming that their distribution
and luminosity is known. For the simplest model of uniformly
distributed, similar sources their number is about several hundreds,
200 -- 400. If we distribute these sources uniformly in the
Universe, the number of sources in the GZK volume with $R=50$ Mpc
would be of the order of $10^{-3}$. This would mean that the
nearest source should be at the redshift $z=0.1$.

A more conservative and self-consistent estimate uses the fact that
protons with energies $E \ge 4\times 10^{19}$~eV observed on Earth can
propagate at most from redshift $z=0.2$ (see, e.g., Fig.~2
in~\cite{photons}). Distributing the sources within a sphere at
$z=0.2$ around the Earth, the closest source is at the distance
$R=100$~Mpc. Note also that in the particular case of BL Lacs as UHECR
sources, which we discuss in Section~\ref{sec:BLLac}, the closest
potential sources are at redshift $z\sim 0.03$.

We show now that the statement that UHECRs with $E\ge 10^{20}$~eV are
protons from nearby sources is in contradiction to the total number of
sources estimated including events below the GZK cutoff. Using Poisson
statistics, the total number of sources $S$ is fixed by the number of
observed singlets $\overline{N}_1$ and doublets $\overline{N}_2$,
\begin{eqnarray}
\label{Poisson1}
\overline{N}_1 &\approx& S \overline{n} \,,\\
\overline{N}_2 &\approx & S \frac{\overline{n}^2}{2} \,,
\label{Poisson2}
\end{eqnarray}
where $\overline{n}$ is the average number of events from a given
source and we assumed equal flux from all sources. From
Eqs.~(\ref{Poisson1}) and (\ref{Poisson2}) one obtains 
the number of sources
\begin{equation}
S \approx \frac{\overline{N}_1^2}{2\overline{N}_2} \,.
\label{Nsimple}
\end{equation}
As it was shown in Ref.~\cite{Dubovsky2000}, the value Eq.~(\ref{Nsimple}) is 
a model independent lower bound on the number of sources for given values of 
$\overline{N}_1$ and $\overline{N}_2$.  In the model of homogeneous
distribution of sources with equal luminosity, the estimate for the
number of sources becomes~\cite{Dubovsky2000}, 
\begin{equation}
S \approx \frac{\overline{N}_{\rm tot}^3}{\overline{N}_{\rm cl}^2} \,,
\label{Nflux}
\end{equation}
where $\overline{N}_{\rm tot}$ is the total number of observed events
and $\overline{N}_{\rm cl}=2\overline{N}_2$ is the number of events in
clusters. In Eq.~(\ref{Nflux}), it is assumed that 
$\overline{N}_{\rm tot} \gg \overline{N}_{\rm cl}$. Note that
therefore Eq.~(\ref{Nflux}) gives always a larger number 
of sources than Eq.~(\ref{Nsimple}) because of the extra factor 
$\overline{N}_{\rm tot}/\overline{N}_{\rm cl} \gg 1$.

We estimate next the number of sources assuming that all UHECR with  $E \ge
4\times 10^{19}$~eV are protons. Following~\cite{Dubovsky2000,FK2000}, we
use 14 events with $E>10^{20}$~eV and one doublet. Calculating $S$ with
Eq.~(\ref{Nsimple}) gives $S \sim 100$ as a minimal number of sources, 
while the more realistic Eq.~(\ref{Nflux}) gives $S \sim 700$. If we 
apply the same analysis to the AGASA data~\cite{AGASAcluster_data2} 
with  $E \ge 4\times 10^{19}$~eV, we have $\overline{N}_2 =6$ 
(we count for simplicity the triplet as one doublet) and $\overline{N}_1=46$. 
Then Eq.~(\ref{Nsimple}) gives  $S \sim 176$, while Eq.~(\ref{Nflux}) gives
$S\sim 1200$. 

Are these two estimates consistent with the idea that all UHECR are
protons? To answer this question, we calculate the expected number of
proton sources using $E\ge 4\times 10^{19}$~eV when there are  $S \sim
100-700$ sources in the GZK volume. Protons with $E \sim 4\times
10^{19}$~eV can reach us from $z=0.2$, or $R_{\rm tot} \sim
1000$~Mpc. Conservatively 
assuming that all events with $E>10^{20}$~eV come from within the GZK
distance $R=50$~Mpc (in \cite{Dubovsky2000,FK2000} $R=25$ Mpc was used),
we obtain with Eq.~(\ref{Nsimple}) as expected number of sources
$S_{\rm tot} = (R_{\rm tot}/R_{\rm GZK})^3 \times 100 = 8\times 10^5$.
Using instead Eq.~(\ref{Nflux}), the expected number of sources is
$S_{\rm tot} = 5.6 \times 10^6$. These estimates should be compared to 
to the ones from AGASA clustering data, $S_{\rm AGASA} \sim 176$ or
$S_{\rm AGASA} \sim 1200 $, using Eq.~(\ref{Nsimple}) or
Eq.~(\ref{Nflux}) respectively.  
Since the Poisson probability to observe $S_{\rm AGASA}$ instead of
$S_{\rm tot}$ events is practically zero, the chance probability to
obtain these two event numbers is equal to the chance probability of
clustering.
We conclude therefore that the model in which {\em all\/} UHECR with  
$E \ge 4\times 10^{19}$~eV are protons from uniformly distributed
point sources is inconsistent with the small scale clustering observed
by AGASA.

One can argue that 14 UHECR events with $E>10^{20}$~eV is an optimistically 
high number and that the real number of such events is much smaller 
because the experiments estimate wrongly the energy of UHECR events.

We conservatively take only the four highest energy events from all
experiments, including one Fly's Eye event, two AGASA events and one
HiRes event. In this case we have 4 single events and no doublets. We
can  estimate the number of sources from  absolute minimal bound 
Eq.~(\ref{Nsimple}) if we assume that the average number of doublets
is less then one, 
i.e. e.g. $\overline{N}_2 =0.5$. Then there are $S=16$ sources in the
GZK volume with $R_{\rm GZK} =50$ Mpc. Again, in a volume with 
$R_{\rm tot} \sim 1000$~Mpc there are $S \sim 128.000$ sources, in
comparison with up to 1200 required by AGASA data above  $E\ge 4\times
10^{19}$~eV. 

Thus, if the clustered component in the AGASA events with energy $E\ge
4\times 10^{19}$~eV  is due to point-like sources, the expected number
of sources is of the order of several hundreds up to $S\sim 1200$,
depending on the estimate used. These sources are distributed in a
volume with $R_{\rm tot}\sim 1000$ Mpc. Assuming that the UHECR events
with $E>10^{20}$ eV are protons requires $10-400$ sources in the GZK
volume with $R_{\rm GZK} = 50$~Mpc. 
These two facts are in contradiction, if all UHECR are protons.
In other words, if UHECR with $E \ge 4\times 10^{19}$~eV  are protons,
we should have less than one source, namely $S\lsim 0.1$, 
in the GZK volume.

\begin{figure}[ht]
\includegraphics[height=0.5\textwidth,clip=true,angle=270]{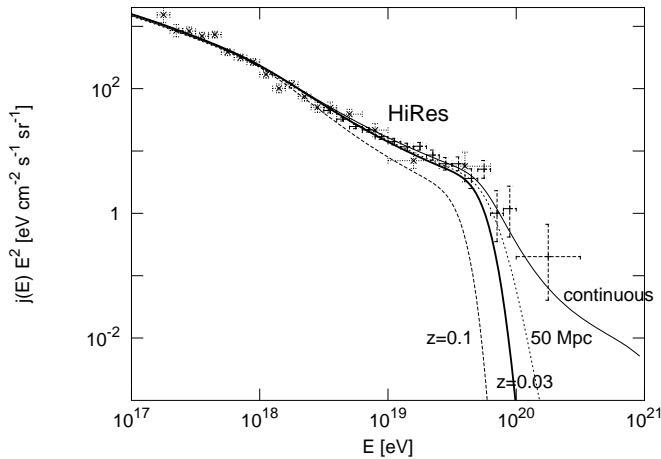}
\caption[...]{UHECR flux measured by the HiRes
  experiment~\cite{Hires}. The thin solid line corresponds to an
  uniform, continuous distribution of proton sources in the Universe
  with emission spectrum $1/E^{2.7}$ and $E_{\rm max} =
  10^{21}$~eV. The dotted curve is for the same model,
  but with no sources within 50 Mpc from the Earth. The thick solid
  line corresponds to no sources within $z_{\rm min}=0.03$, the dashed
  line to $z_{\rm min}=0.1$.} 
\label{F01}
\end{figure}

\begin{figure}[ht]
\includegraphics[height=0.5\textwidth,clip=true,angle=270]{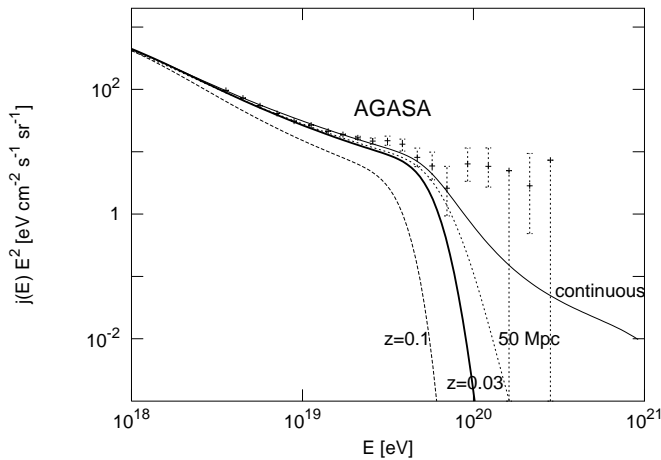}
\caption[...]{UHECR flux measured by the AGASA
  experiment~\cite{agasa}. All other parameters are the same as in
  Fig.~\ref{F01}.} 
\label{F02}
\end{figure}

Let us now discuss the consequences of a small number of sources
for the model of uniformly, continuously distributed point sources of
protons. For our calculations, we have used the code developed in
Ref.~\cite{code2}, in which all important effects (pion production,
$e^+e^-$ production and the expansion of the Universe) are taken into
account.
Essentially, we have repeated for the case of proton primaries and BL
Lacs as sources the calculations made in~\cite{photons} for photons in
more details. In Figs.~\ref{F01} and \ref{F02} we show with thin solid
lines the spectrum of continuously distributed sources of protons with emission
spectrum $1/E^{2.7}$ and $E_{\rm max} = 10^{21}$eV as in~Ref.~\cite{BGG2002}.
The dotted, thick solid and dashed lines are for the same model, but
with no sources within  50 Mpc, $z_{\rm min}=0.03$ and $z_{\rm min}=0.1$
around the Earth, respectively. The minimum distance of $z_{\rm min}=0.03$
corresponds to the BL Lac distribution.

Let us concentrate on Fig.~\ref{F01}, which shows the measured
spectrum of HiRes and where the fit model of~\cite{BGG2002} with an
infinite number of sources (thin solid line) works well. If there are
no sources within 50~Mpc (dotted curve), the two highest HiRes data
points are well above the model fit. For the  BL Lac case where the
closest known sources are at $z_{\rm min}=0.03$ two additional
experimental points are away from the fit. Finally, for an
uniform distribution of 400 proton sources over the Universe, or
$z_{\rm min}=0.1$, the disagreement above the cutoff becomes even worse.
Note that we are only concerned about energies above $\sim 6\times
10^{19}$~eV; at lower energies, the quality of  the fitted model can
be easily improved by a readjustment of the fit parameters.
The same figure with experimental data from AGASA is shown
in~Fig.~\ref{F02}. 

Thus, if the clustered component of the AGASA data for $E \ge 4\times
10^{19}$~eV (which has a statistical significance of $4.6\sigma$) is
not a statistical fluctuation or the result of magnetic lensing, the
expected relative small number of UHECR sources is inconsistent with
the model of proton primaries emitted by uniformly continuously
distributed sources 
both for the HiRes and AGASA data. This means that {\em both\/} the
AGASA and HiRes data require the introduction of a new component (not
protons) in the UHECR spectrum. In the following sections we will
consider new light hadrons with the mass of 2--3 GeV as such a new
component.


\section{Light strongly interacting particles: models and experimental
        status} 
\label{models}

We shall concentrate our discussion on the definite case of the light
sbottom, which is rather predictive and allows various tests of our
assumptions. 
Since an useful UHE primary should be stable or quasi-stable with
life-time $\tau\gsim 1$~month (cf. Sec.~\ref{propagation}), only the
LSP or the NLSP are possible
candidates as new UHE messengers in SUSY models with (approximately)
conserved R parity. The NLSP as UHE primary  can be realized if it
has a very small mass splitting with the LSP or if the LSP is the
gravitino; in the latter case the gluino decays gravitationally and its
lifetime can be long enough.  

Theoretically the best motivated candidates for the LSP are the
neutralino $\tilde\chi$ and the gravitino $\tilde G_{3/2}$. 
While in minimal supergravity models the
LSP is the lightest neutralino (in some part of the parameter space it
is the sneutrino), in models with gauge-mediated SUSY the LSP is
normally the gravitino.
 Recently, a light bottom squark $\tilde b$ with
mass $m_{\tilde b}\sim 2-6$~GeV has been suggested~\cite{b}, motivated
by the large bottom quark production cross section measured at the
Tevatron~\cite{Tevatron}: The long-standing puzzle of overproduction
of $b\bar b$ pairs can be solved if there exists additionally to a
light sbottom a light gluino with mass $m_{\tilde g}\sim  12-16$~GeV.

Bottom squarks as LSP can either form charged  
$\tilde B^-=(\tilde b \bar u)$ or neutral $\tilde B^0=(\tilde{\bar b} d)$ 
(plus charged conjugated) two-quark states. Since $qu$ states
are generally lighter than $qd$ states, it is likely that the
charged $\tilde B^\pm=(\tilde b  u)$ is lighter than the neutral 
$\tilde B^0=(\tilde b d)$. But their mass difference will be very
small, e.g., for the usual $B$ system $m_{B^0}-m_{B^\pm}\sim 0.33\pm
0.28$~MeV, and we consider therefore the question if the lightest
state would be charged or neutral as open. Moreover, the mass
difference in the $\tilde B$ system could be smaller than the electron
mass, and weak decays therefore kinematically forbidden. In this case,
both the $\tilde B^\pm$ and the $\tilde B^0$ would be stable.
Apart from the two-quark states, there will be again baryonic
three-quark states, like e.g. $\tilde b ud$. These baryonic
states can decay into a baryon and a $\tilde B$ if kinematically
possible.
 
Theoretically, the lightest
hadronic state could be electrically charged. Is it possible that a
light, stable charged hadron evaded detection? At Serpukov and the
CERN ISR several searches for such particles were performed
in the 70's~\cite{sep,chlm,br-scan}.   
For example, the CHLM experiment excluded the range $m\geq
2.4$~GeV for stable hadrons with charge $q=1$~\cite{chlm}. Below
2.4~GeV, the production of antideuteron could hide other hadrons with
a similar mass. Since the ratio $R$ of antideuteron to pion
production in these experiments is rather high, $R\sim 5\times
10^{-4}$~\cite{br-scan}, and 
the mass resolution of these experiments not too fine, a significant  
fraction of deuterons could be misidentified stable charged hadrons. 
Also the TRISTAN experiment did not include the deuteron region in
their search for massive stable hadrons~\cite{topaz}.
While the LEP experiments, in particular DELPHI, could exclude
generally charged shadrons down to masses of 2~GeV, the limit is in
the case of a sbottom with small couplings to the $Z$ boson weakened and
sbottoms with masses below 5~GeV are allowed~\cite{DELPHI}.
The ALEPH exclusion limit was not extended to masses below
5~GeV~\cite{aleph}.   
The CLEO experiment was able to exclude charged hadrons with mass
$m\leq 3.5$~GeV, however only for fractionally charges~\cite{cleo}.
Since there is however also no positive evidence for a stable charged
hadron, we consider mainly the option that the lightest sbottom
containing hadron is neutral.

Next we recall that a light sbottom quark is consistent with
electroweak precision observable and with the LEP higgs mass
limit~\cite{Carena:2000ka}. Reference~\cite{Cho:2002mt}
showed that this scenario implies a light stop, $m_{\tilde t_1}\lsim
98$~GeV, offering the TEVATRON run-II experiments the possibility to
(dis-) prove indirectly the light sbottom case. 
The observation of a $\bar{\tilde b}\tilde
b$ resonance in $e^+e^-$ annihilation is difficult to extract from the
background because the $\bar{\tilde b}\tilde b$ resonance has to be
produced in a p-wave~\cite{Nappi:1981ft}. 
Its contribution to $e^+e^-\to$~hadrons is small
compared to the error of these measurements. Since the lightest $\tilde B$
behave as a stable particle in any accelerator experiment, their
identification would require a dedicated analysis. We consider
therefore a sbottom with mass  1.5--3~GeV 
as a viable option and shall investigate its use as a UHE
primary in the subsequent sections.

Finally, we discuss the case of rather short-lived shadrons.
Possible decays are $\tilde b\to b+\tilde G_{3/2}$ in models
where the gravitino is the LSP or decays like $\tilde G\to\pi+\nu$,
etc., if $R$-parity is violated. In Ref.~\cite{OV}, it was argued
that these decays can be excluded by proton decay 
experiments. If the life-time of $\tilde g$ or $\tilde b$ is close to
the required lower limit of $\sim 1$~month, the shadrons produced by
cosmic rays and contained in the detector material have time to decay
during the build-up and start phase of the experiment. Since these
experiments are deep underground, they are well shielded and cosmic
rays or shadrons cannot reach the detectors. In the case of
detector using purified detector material, originally contained
shadrons could be also extracted in the purification
process, depending on the chemical properties of the shadrons.  

Light sbottoms do not contradict cosmological
limits: The relative abundance of gluinos is $n_{\tilde g}/n_\gamma\sim
10^{-20}- 10^{-17} (m_{\tilde g}/{\rm GeV})$ \cite{Ba99} and
possible decays do not disturb Big Bang
Nucleosynthesis (BBN)~\cite{cyburt}. If baryon number also resides in
baryons $\tilde N$ containing sbottoms, then their number is
suppressed by $n_{\tilde N}/n_B\sim \exp(-M_{\tilde N}/T_{\rm QCD})\sim
\exp(-3/0.16)\sim 10^{-7}$, where $T_{\rm QCD}$ is the temperature
of the QCD phase transition. Again there is no conflict with BBN.

We should stress that light sbottoms in SUSY theories 
serve as merely as examples. Any  particle physics model which has a
(quasi-) stable particle with mass 1.5--3~GeV and interacts strongly
with protons should have similar consequences for the physics of UHECR.

\section{Propagation through the Universe: how to avoid GZK cutoff}
\label{propagation}

Strongly interacting particles $S$ propagating through the Universe
interact with CMB photons producing pions, if their energy is above
the single pion production threshold, 
\begin{equation}
E_{\rm th} = \frac{m_\pi^2 + 2 m_\pi M_S}{4\epsilon_0} \,.  
\label{GZKcutt}
\end{equation}
Here, $m_\pi$ and $M_S$ denotes the mass of a pion and of $S$,
respectively. For the following simple estimates, we neglect  
the Bose-Einstein distribution of the photon energies and use just the
average energy of a CMB photon, 
$\epsilon_0 = \pi^4 T_0/30 \zeta (3) \approx 6.4 \times 10^{-4}$ eV.
In order to avoid the GZK cutoff, the cross section of strongly
interacting hadrons $S$ with photons should be smaller than the one of
nucleons, 
\begin{equation}
\sigma_{S\gamma} = B \left(\frac{m_p}{M_S}\right)^2 \sigma_{p\gamma}~,
\label{GZKcross}
\end{equation}
where the suppression factor in the resonant case comes from the
different center mass momenta in the Breit-Wigner formula.
The dimensionless parameter $B$ is $B(s)\lsim 1$, because the
assumed resonant state (the equivalent of the $\Delta$ resonance)
has a mass larger than $S$.

Thus, strongly interacting UHE particles with $E\gg E_{\rm th}$ will
interact with CMB photons on the typical scale 
\begin{equation}
l_{\rm int} = \frac{1}{\sigma_{S\gamma} n_0} = 
8~{\rm Mpc}\; \frac{\sigma_{p\gamma}}{\sigma_{S\gamma}} \,,
\label{GZKlength}
\end{equation}
where the CMB number density is $n_0 =410$ cm$^{-3}$ and
$\sigma_{p\gamma}=10^{-28} {\rm cm}^2$  is the multi-pion production 
cross section. During each interaction, the particle $S$ loses the
fraction $y\sim 0.5$ of its energy until its energy is close to
$E_{\rm th}$. There, the energy fraction lost reduces to $m_\pi/M_S$,
while the cross section can be increased due to resonances.

If one defines the radius $R_{\rm GZK}$ as the distance after which a
particle $S$ with $E\gg E_{\rm th}$ lost 95\% of its initial energy,
then 
\be
 R_{\rm GZK} = \frac{\ln 20}{y\sigma_{S\gamma}n_\gamma} 
 \sim 48 \, \frac{\sigma_{p\gamma}}{\sigma_{S\gamma}} {\rm Mpc} \,. 
\ee
In the case of protons, this effect was first considered by Greisen,
Zatsepin and Kuzmin in 1966 \cite{gzk} and is called GZK effect. The
threshold energy for proton from Eq.~(\ref{GZKcutt}) is $E = 1 \times
10^{20}$ eV. However, protons with energies $E_{\rm GZK} = 4 \times
10^{19}$ eV can still interact with the high-energy tail of the CMB
distribution.   

\begin{figure}[ht]
\includegraphics[height=0.48\textwidth,clip=true,angle=270]{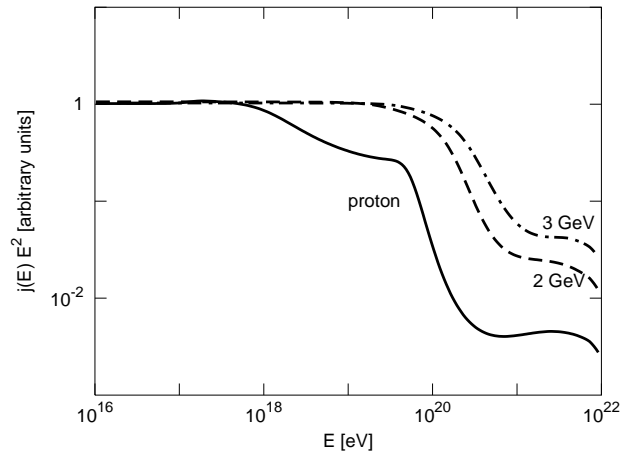}
\caption[...]{
Energy spectrum  of $\tilde B^0$ hadrons with injection spectrum
$E^{-2}$ and uniformly distributed sources for a  
$M_{\tilde B}= 2$ and 3~GeV; for a comparison proton spectrum is also shown.
At energies $10^{18} ~{\rm eV}~< E < 6 \times 10^{19} ~{\rm eV}$
proton spectrum also  
suppressed due to $e^+e^-$ production.
\label{F0}}
\end{figure}

Let us now consider the case of a new, strongly interacting particle
$S$ from a general point of view. If its heavier than a proton, then
its GZK cutoff is both softened and shifted to higher energies.
The first effect arises because of the smaller energy transfer near
threshold, while the second one is due to a smaller resonant 
cross section with CMB photons.
Let us now turn to our specific examples $\tilde G=(\tilde gg)$, and
$\tilde B^0=(\tilde b d)$ and $\tilde B^\pm=(\tilde b u)$. The first
case was studied in 
Ref.~\cite{Berezinsky:2001fy}. It was found that $\sigma_{\tilde
  G\gamma}$ is at 
least a factor 8 smaller than $\sigma_{p\gamma}$, even for as low
masses as $m_{\tilde g}=1.5$~GeV. This small cross section leads
together with the reduced energy losses per scattering to a shift of
the GZK cutoff close to the maximal energies of astrophysical
accelerators~\cite{Berezinsky:2001fy}. 
While in the case of $\tilde G$ some information about the
mass spectrum of low-lying $\tilde g$ containing hadrons is
available~\cite{Chanowitz:1983ci}, this information is missing for  
$\tilde B^{0,\pm}$. Since the knowledge of the low-lying
resonances of $\tilde B$ is essential to perform a
detailed calculation of its energy losses on CMB photons, we can only
estimate the energy losses. To be conservative, we assume that the
resonant contribution to $\tilde B\gamma$ scattering is only
suppressed by its larger mass, or $B=1$ in Eq.~(\ref{GZKcross}), and
by the smaller energy transfer close to 
threshold, $y=m_\pi/M_{\tilde B}$. Using as smallest value for
$M_{\tilde B}\sim 2$~GeV then shifts $E_{\rm th}$ by a factor two, what
together with the other suppression factors causes only a mild GZK
effect at high enough energies. The resulting spectrum is shown for
$\tilde B^0$, an injection spectrum $E^{-2}$ and uniformly
distributed sources for a $M_{\tilde B}= 2$ and 3~GeV in
Fig.~\ref{F0}. In the case of the charged $\tilde B^\pm$, additionally
$e^+e^-$ pair production has to be considered. For comparison we show
in Fig.~\ref{F0} the proton spectrum from the same distribution of
sources.   

Another important condition is that the particle $S$ should be stable,
traveling through the Universe. 
The lifetime $t_S$ should be bigger then
\begin{equation}
t_S = \frac{R_U}{c} \frac{M_S}{E_{\rm UHE}} \approx 12 ~{\rm days}~~
\frac{R_U({\rm Gpc}) M_S({\rm GeV})}{E_{20}}~, 
\label{GZK_decay}
\end{equation}
where $R$ is measured in Gpc, $M_S$ in GeV and $E_{20} = E/(10^{20}
{\rm eV})$. Note that this allows the possibility that the gravitino
is the LSP and the gluino or bottom squark is the NLSP. 
In this case, the NLSP decays only via
gravitational interaction and thus has a long enough life-time to
serve as UHECR messenger. On the other hand, all experimental
constraints from searches for anomalous heavy isotopes can be avoided
easily in this scenario. 

Thus any new strongly interacting ``messenger'' particle with
multi-GeV mass and lifetime bigger than a year can travel over
cosmological  distance and solve the GZK problem.    
In particular, gluino and bottom squark containing mesons and baryons can 
serve as as ``messenger'' particles.

\section{Interactions in the atmosphere}
\label{atmosphere}

The interactions of glueballinos with nucleons were considered in
detail in Ref.~\cite{Berezinsky:2001fy}. There, the Monte Carlo
simulation QGSJET~\cite{QGSJET}, which describes hadron-hadron
interactions using the quark-gluon string model of the supercritical
Pomeron in the framework of the Gribov-Regge approach, was extended to
include light gluinos. Moreover, extensive air showers (EAS) were
simulated and the resulting 
lateral and longitudinal shower profiles were compared to those of EAS
initiated by protons. The authors of Ref.~\cite{Berezinsky:2001fy}
concluded that glueballinos with mass $\gsim 5$~GeV resemble a penetrating
particle and  can be already excluded using existing data, while 
EAS initiated by glueballinos with mass $\lsim 3$~GeV can be only
differentiated from proton showers by future experiments with larger
statistics. 

The calculations of Ref.~\cite{Berezinsky:2001fy} were done only for
the special case of a glueballino. However, $\tilde B$-hadrons with
the same mass should have very similar interaction properties.
Main reason for this is that the coupling of the Pomeron to a hadron
as well as the slope of its coupling depends essentially on the size
of the hadron, and therefore on its reduced mass. Minor differences
arise because of the different constituent masses of quarks and gluons
resulting in different momentum distributions of gluino and squarks in
different hadrons with the same mass. Otherwise, the soft and semi-hard
interactions have the same dependence on its mass.
Finally, the hard interactions of the constituents at UHE
energies are practically mass independent in the low mass
range of interest.  
We conclude therefore that also
$\tilde B$-hadrons with mass $\lsim 3$~GeV produce EAS consistent with
present observations.

\section{Production of SUSY particles in astrophysical sources}
\label{sources}

Protons are the most natural candidates for the observed UHECR with
energies above the ankle $E>10^{19}$~eV. There are several mechanisms
which could be responsible for the acceleration of protons to the
highest energies: The most popular one is particle acceleration in
shock fronts or Fermi acceleration of the first kind. 
However, there are other, more exotic mechanisms as, e.g.,  
particle acceleration in the vicinity of black holes rotating in an 
external magnetic field (see for example~\cite{neron}). 

Independent of the specific acceleration mechanism, a simple estimate   
of the maximal possible energy up to which a source can accelerate
particles was suggested by Hillas~\cite{hillas}. It is based on the
relation $E_{\rm max} = qBL$, where $q$ is the charge of the 
accelerated particle, $B$ the magnetic field strength in the
acceleration region and $L$ its size. Only few astrophysical objects
are able to accelerate particles to UHE according to this simple criterion. 
Plausible candidates for acceleration to UHE are AGNs and several
AGN subclasses were suggested as sources of UHECRs.
The general perception is that it is possible to accelerate protons in
objects like AGNs up to $E_{\rm max} \lsim 10^{21}$ eV, but that 
acceleration to higher energies is extremely difficult because of
energy losses.

\subsection{Proton-proton interactions}

We start with the perturbative calculation of the production cross
section of bottom squarks in proton-proton collisions.
The main contribution to the total cross-section is given by the
gluon-gluon subprocess, first calculated at leading order in
\cite{old}. The center-of-mass energy $\sqrt{s}$ of this process is
rather high, $\sqrt{s}\gsim 300$~TeV, and we restrict ourselves
therefore to a leading-order calculation. We have used the Cteq6 
parton distribution functions (pdf)~\cite{cteq6} with the scale $\mu^2=\hat s$
and calculated as main contribution to the total production cross
section the parton subprocesses $gg\to \tilde g\tilde g$  and
$gg\to \tilde b\overline{\tilde b}$. 
For the relatively large gluino masses still allowed, $m_{\tilde
    g}\gsim 6$~GeV, the cross section of $gg\to \tilde g\tilde g$ is
  too small and we do not discuss this case in the following.

\begin{figure}[ht]
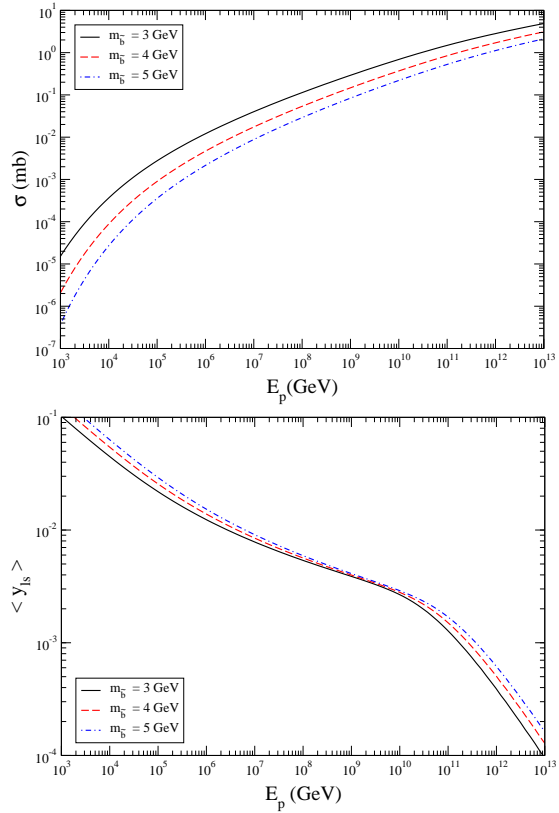

\includegraphics[height=0.3\textwidth,clip=true,angle=0]{2.cross_new.eps}
\includegraphics[height=0.3\textwidth,clip=true,angle=0]{2.xls_new.eps}
\caption[...]{(a) Bottom squarks production cross section in
  proton-proton interaction as function 
of UHE proton energy. (b) Energy fraction transfered to bottom squarks
from initial protons as function of UHE proton energy}
\label{F2ab}
\end{figure}

The sbottom production cross section  as function of the UHE proton
energy $E_p$ in the lab frame is shown in Fig.~\ref{F2ab}a for masses
$M_{\tilde b} =3,4$ and 5~GeV. At energies $E_p\sim 10^{13}$~GeV,
this cross section reaches several mbarn.
The fast increase of the cross section with energy is caused by the
growing number of accessible soft gluons with smaller and smaller
$x_{\rm min}(s)$ values. Therefore, the prize to be paid for such a 
large cross section is the small energy fraction transferred,  
$\langle y_{\rm ls}\rangle = 
\langle E_{\tilde b}/E_p\rangle \sim 10^{-3}$ for $E_p=10^{20}$~eV, 
see Fig.~\ref{F2ab}b. Such small values of $y$ make it impossible to
produce UHE sbottoms in pp-collisions: even if the primary protons
would have energy $E_p=10^{23}$~eV, the average energy of the produced
sbottom would be only $10^{19}$~eV. Since $10^{4}$ more energy will be
dumped into neutrinos and photons than into sbottoms, it is impossible
to explain the UHECR flux $F_{CR}$,
\begin{equation}
F_{CR}(E)  = \left(\frac{10^{20} {\rm eV}}{E}\right)^2 
             \frac{{\rm eV}}{{\rm cm}^2 {\rm s\,sr}} \,,
\label{UHE_flux}
\end{equation}
with sbottoms without overproducing photons and neutrinos. 
The produced photons will cascade down to GeV energies and
overshot the diffuse gamma-ray flux measured by EGRET~\cite{egret} by
two orders of magnitude. In principle, one can argue, that it is
possible to transfer this energy already in the source to energies
below those measured by EGRET energies, thereby avoiding the 
EGRET bound. However, at the same time the neutrino flux of the order
of $10^{5}$~eV/(cm$^2\; {\rm s}\; {\rm sr})$ will overshut 
existing limits on neutrino flux, given by Fly's Eye
\cite{Fly'sEye_nu}, AGASA~\cite{agasa_nu} and RICE~\cite{rice}.

\subsection{Proton-photon interactions}

\begin{figure}[ht]
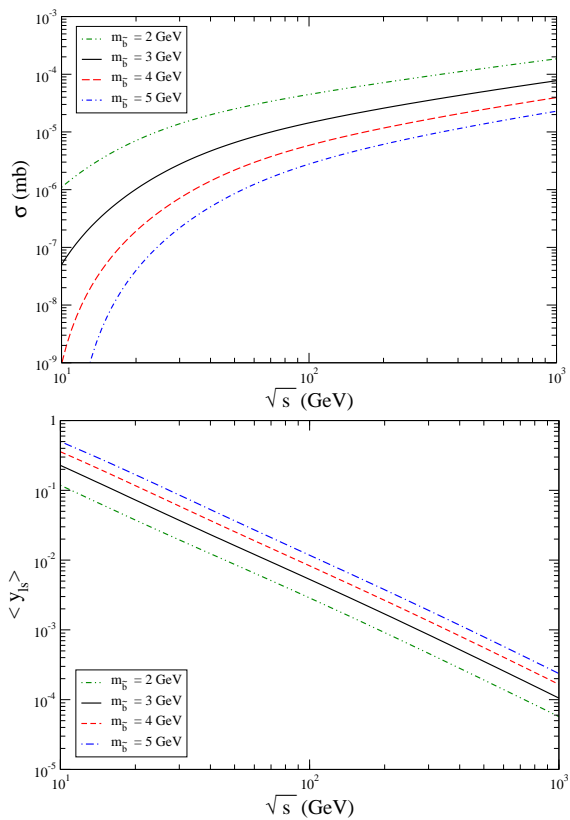

\includegraphics[height=0.3\textwidth,clip=true,angle=0]{3.cross_new2.eps}
\includegraphics[height=0.3\textwidth,clip=true,angle=0]{3.xls_new2.eps}
\caption[...]{(a) Perturbative part of bottom squarks production cross 
section in proton-photon interaction as function
of center mass energy. (b) Part of the energy transfered to bottom squarks from initial protons 
as function
of center mass energy in perturbation theory. }
\label{F3ab}
\end{figure}

We consider next the perturbative calculation of the production cross
section of bottom squarks in proton-photon collisions.
Now, the most important subprocesses for squark production is the
parton subprocess $\gamma g\to \tilde b\overline{\tilde b}$.  
The cross section for sbottom production
is shown in Fig.~\ref{F3ab}a. 
Even compared to the relatively small total $p\gamma$ cross section
which is of the order of $0.1$~mb this cross section is small.
However, now the center-of-mass energy can be much smaller: the typical 
energy of an infrared/optical target photon is in the range 0.1-10~eV,
hence $\sqrt{s}$ is between the production threshold and several
100~GeV. Therefore, the energy fraction transferred is now much
higher, see Fig.~\ref{F3ab}b. However, the combination of these
two suppression factors is again very small. Normalizing the UHE
$\tilde g$ or $\tilde b$ flux to the UHECR flux, Eq.~(\ref{UHE_flux}), 
would produce $10^6$ times higher neutrino and photon fluxes, which is
again in contradiction both with diffuse gamma-ray flux measurements
and with neutrino bounds.

Apart from the perturbative contribution to the cross section
calculated above, non-perturbative contributions have to be considered
where at small momentum transfer hadrons interact with each other. 
A calculation of this kind could be performed in the
vector dominance model. Then the total production cross section can be
split into two parts, 
\begin{equation}
 \sigma_{p\gamma}^S \approx 
 \sum_i \alpha w_i \sigma_{ip} +
 \frac{3\alpha e_q^2}{2\pi} 
 \int_{k_\perp^2>Q_0^2} \frac{dk_\perp^2}{m_{\tilde b}^2+k_\perp^2} \:
 \sigma_{\gamma\to qq}(s,k_\perp^2)
\end{equation}
where the sum $i$ extends over the vector mesons $i$ with weight $w_i$
and the second part describes the perturbative process 
$\gamma\to\bar{\tilde b}\tilde b$ matched to the first contribution
at  $Q^2=Q_0^2$~\cite{ralph}. 
The second contribution can be evaluated at UHE and gives 
$\sigma_{p\gamma}^S \sim (m_\pi/m_{\tilde b})^2 \sigma_{p\gamma}$.
The dominant subprocess of the first part is the $t$ channel
exchange of $\tilde B$ mesons. It is therefore natural to expect that
this contribution is also suppressed relative to the photon-proton total
cross-section in the multi-pion production region by the ratio
$m_\pi^2/M_S^2$. We shall parameterize therefore the bottom squark
production cross section as 
\begin{equation}
\sigma_{p\gamma}^S = A \frac{m_\pi^2}{M_S^2} \sigma_{p\gamma} \,,
\label{cross_section}
\end{equation}
where $A(s) \lsim 1$ is a dimensionless factor depending on $s$. We expect
$A(s)\sim 1$ in the high-energy region and $A(s)\to 0$ for $s\to
4m_{\tilde b}^2$. The transferred energy
can be as high as $10-50$\%. The required photon energies are of the
order of $0.1-10$ eV. 

Since we are interested in shadron masses around 1.5--3~GeV, we can
compare the value predicted by Eq.~(\ref{cross_section}) with the
total charm production cross section in proton-photon collisions. 
Reference~\cite{Berezhnoy:ji} collected experimental
data~\cite{photo-production-exp} in the energy range from
$\sqrt{s}=10$--200~GeV; the cross section increases from
$10^{-3}$~mbarn at $\sqrt{s}=20$~GeV to $10^{-2}$~mbarn at
$\sqrt{s}=200$~GeV. According to (\ref{cross_section}), 
we would expect a cross section 
$(m_\pi/M_D)^2 \sigma_{p\gamma}\sim 5\times 10^{-3}
\sigma_{p\gamma}\sim 10^{-3}$~mbarn. Hence we conclude that 
Eq.~(\ref{cross_section}) is a rather conservative estimate.

Close to threshold, $t$ channel exchange of $\tilde B$ mesons proceed
as $p+\gamma\to (ud\tilde b)+(\bar{\tilde b}u)$ and 
$p+\gamma\to (uu\tilde b)+(\bar{\tilde u}d)$. Thus at moderate UHE
energies, the UHECR flux should consist of the usual protons, and
positive and neutral $\tilde b$-hadrons. At the highest energies,
when several $\tilde b$-hadrons are produced, additionally negatively  
charged $\tilde b$-hadrons appear.

Let us compare these numbers with the parameters of astrophysical objects.  
The important difference to proton-proton interactions is that the
energy of background photons and as a result the center-of-mass energy
is normally much smaller than the proton mass. The required
center-of-mass energy to produce particles with mass in the 
multi-GeV range is around $s=50~ {\rm GeV}^2 s_{50}$. It should be
somewhat higher than the threshold $4M^2$ to avoid the kinematical
suppression effects near threshold. The typical photon energy then is
\begin{equation}
\epsilon = \frac{s}{2 E_p} = 0.25 \frac{s_{50}}{E_{20}} ~ {\rm eV}\,.
\label{photon_energy}
\end{equation}

Photons of such energies exist in many astrophysical objects which can
accelerate protons.  However, the accelerated protons should interact
inside these objects with photons. In other words, the propagation
length $l_{\rm int}$ of protons should be smaller than the size $R$ of
the interaction region,  
\begin{equation}
l_{\rm int} = \frac{1}{\sigma_{p\gamma} n_{\gamma}}< R \,.
\label{size}
\end{equation}
For example, the time variability of several days in the optical
spectrum of AGN cores corresponds to a region size of $R=10^{16}$
cm \cite{opt_var}. The photon density can be estimated from the
optical/infrared luminosity, 
\begin{equation}
n_\gamma = \frac{L}{4\pi c R^2\epsilon_\gamma} = 10^{13}
\frac{1}{\rm cm^3} \frac{L_{44}}{R_{16}^2 \epsilon_{-1}}~, 
\label{n_photon}
\end{equation}
where the quantities introduced are the dimensionless luminosity
$L_{44}=L/(10^{44}{\rm erg}/{\rm s})$, the region size
$R_{16}=R/(10^{16}$~cm) and the typical photon energy $\epsilon_{-1} =
\epsilon/(0.1$~eV).  
Substituting the multi-pion production cross section $\sigma_{p
  \gamma} \approx 10^{-28} {\rm cm}^2$ and  
Eq.~(\ref{n_photon}) into the condition (\ref{size}) we obtain
\begin{equation}
\tau = \frac{R}{l_{\rm int}} = 10 \frac{L_{44}}{R_{16} \epsilon_{-1}} \gg 1~.
\label{depth}
\end{equation}
Thus, if the parameters of the source are similar to those in
Eq.~(\ref{depth}), protons produced inside such sources will interact
with background photons and can potentially produce secondary hadrons
$S$. However, the produced hadrons still need to escape
from the astrophysical source. The escape condition is inverse to the 
one given in Eq.~(\ref{size}): The optical depth for the new hadrons
should be small assuming the same source parameters.  Then the escape
condition is 
\begin{equation}
\sigma_{S\gamma} < \frac{1}{\tau_{p\gamma}} \sigma_{p\gamma}~,
\label{escape}
\end{equation}
where the optical depth for protons $\tau_{p\gamma}$ is defined  by
Eq.~(\ref{depth}).
In the case of a light glueballino, the suppression of the
$\sigma_{S\gamma}$ cross section can be of the order of 0.1 in
comparison to $\sigma_{p\gamma}$ at center mass energies of the order
of 10~GeV~\cite{Berezinsky:2001fy}. This is consistent with
Eq.~(\ref{escape}), if the parameters of the astrophysical object are
the same as in Eq.~(\ref{depth}). 
In the case of a light bottom squark we suppose that 0.1 is also a
reasonable estimate for the suppression factor in the
$\sigma_{S\gamma}$ cross section and leave the detailed analysis for
future investigation.

Thus, new light hadrons can be produced in astrophysical objects from
$10^{21}$~eV protons, interacting with infrared/optical photons of the
energies $0.1-10$~eV, if the sources are
optically thick for protons, Eq.~(\ref{depth}).
The model for such a source can be similar, for example, to  the one
of Stecker {\it et al.}~\cite{stecker}. 
Produced hadrons will escape from the same objects if their interactions with
photons are suppressed in comparison to those of protons,
Eq.~(\ref{escape}). However, simultaneously with the new hadrons 
large fluxes of neutrinos and photons will be produced unavoidably. 
In next section we will discuss the experimental constraints on these
fluxes.

We have used AGN cores as a working example of astrophysical accelerators, 
which, as we have shown, can obey the condition of high optical depth for
protons, Eq.~(\ref{depth}), and allow shadrons to escape,
Eq.~(\ref{escape}).  Any other astrophysical object, which is able to 
accelerate protons to the highest energies and obeys these 
conditions can be a source of  shadrons as well.

\section{Allowed parameters of new hadrons consistent with gamma-ray and
neutrino bounds}
\label{parameters}

\begin{figure}[ht]
\includegraphics[height=0.5\textwidth,clip=true,angle=270]{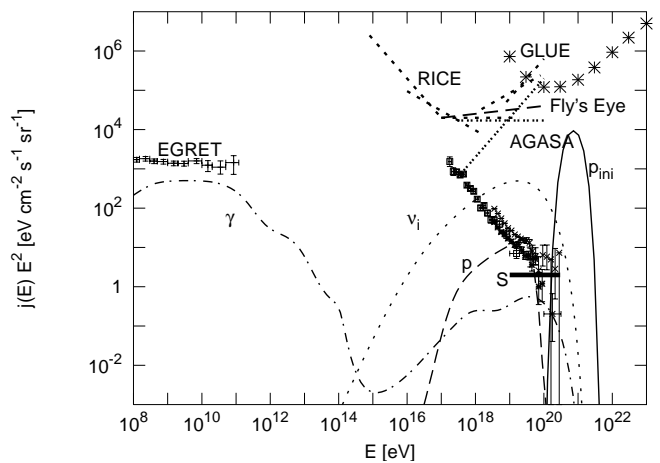}
\caption[...]{Flux of new hadrons $S$ (thick solid line) and protons
  (dashed line) together with cosmic ray data from AGASA~\cite{agasa}
  and HiRes~\cite{Hires}. 
  Protons accelerated to the energy $E=10^{21}$ eV (line $p_{\rm
  ini}$) produce secondary photons (dashed-dotted line) and neutrinos
  (dotted line). Photon flux constraint from EGRET~\cite{egret} and
  upper limits on the diffuse neutrino fluxes from
  AGASA~\cite{agasa_nu}, the Fly's Eye~\cite{Fly'sEye_nu}, the
  RICE~\cite{rice}, and the Goldstone experiment (GLUE)~\cite{glue}
  as indicated.}
\label{F4}
\end{figure}

\begin{figure}[ht]
\includegraphics[height=0.5\textwidth,clip=true,angle=270]{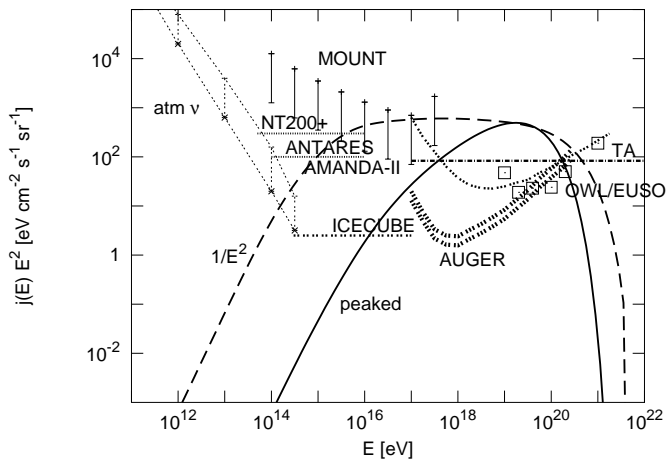}
\caption[...]{The neutrino flux for one flavor in the model used in
  Fig.~\ref{F4} and sensitivities of the currently being constructed
  Auger project to electron/muon and tau-neutrinos~\cite{auger_nu}, 
  and the planned projects telescope array (TA)~\cite{ta_nu}
  (dashed-dotted line), MOUNT~\cite{mount}, and, indicated by squares, 
  OWL~\cite{owl_nu}, NT200+~\cite{baikal_limit},
  ANTARES~\cite{antares}, AMANDA-II and ICECUBE~\cite{icecube}, as
  indicated. Also shown (dashed line) is an extreme scenario with
  initial proton spectrum $1/E^2$, leading to a neutrino flux
  extending to relatively low energies where Baikal, ANTARES and
  AMANDA-II will be sensitive, and the atmospheric neutrino flux for
  comparison.} 
\label{F5}
\end{figure}

As it was shown in the previous section, only interactions of UHE
protons with infrared/optical background photons of the energy $0.1-10$ eV
can produce a significant amount of new strongly interacting hadrons
$S$, without overproducing the diffuse photon and neutrino
backgrounds. The essential condition for this mechanism is the high
optical depth for protons, Eq.~(\ref{depth}). However, this condition
has not to be fulfilled by all UHE proton sources. Sources with small
optical depth for protons will just emit UHE protons, which will be
responsible for UHECR below the GZK cutoff. The few sources of UHE
shadrons will be only responsible for the highest energy cosmic rays
with $E>10^{20}$~eV and may be, partly, for the clustered component at
lower energies. It is also not excluded that few sources of this kind 
will responsible for most, or all UHECR above $4 \times 10^{19}$~eV. 

We do not specify the proton acceleration mechanism in the
astrophysical objects here. We just suppose that these sources can
accelerate protons up to $10^{21}$~eV. The proton spectrum should be
relatively hard in this case, $1/E^\alpha$ with $\alpha \le 2$. The
exact shape of the spectrum is not important, because the optical
depth for protons is high and most of them are absorbed. The pion
production threshold for an optical background of 10~eV is $E_{\rm th}
= 2 \times 10^{16}$~eV. Thus the flux of protons with energy above this
threshold will be reduced by the factor $\exp(-\tau(E))$. If the
optical depth is high, $\tau (E) \gg 1$, all protons 
will be absorbed in the source, and the proton flux does not overshoot
the measured spectrum of UHECRs.

We have used the propagation code~\cite{code2} for the calculation of
the energy spectra of secondary protons,  photons and neutrinos. As
initial spectrum, we have chosen a proton spectrum peaked at the energy
$10^{21}$~eV, see Fig.~\ref{F4}. The continuation of this spectrum to
lower energies is possible for any power law up to $\alpha \le 2$. 
Even an initial proton spectrum with $1/E^2$ will not contradict UHECR
observations at high energies, but only will lead to a higher flux of
UHE neutrinos at the energies  $10^{16}{\rm eV}\le E \le 10^{20}$~eV.   

In Fig.~\ref{F4} we present one example of such a calculation. Cosmic
ray data from  AGASA~\cite{agasa} and HiRes~\cite{Hires} are shown
with errorbars. The contribution of the new hadrons $S$ to the UHECR
spectrum at the highest energies is shown with a thick solid line. We
have used conservatively the AGASA spectrum at the highest energies as
normalization -- choosing HiRes data as reference would increase the
parameter space for the new hadron. The exact shape of the $S$ particle
spectrum is not well defined, because it depends on many unknown
parameters like the spectrum of background photons in the source, the 
distribution of the sources, initial proton spectrum and the energy
dependence of the production cross section. However, the amplitude of
this flux is related to the amplitude of the initial proton flux through
Eq.~(\ref{cross_section}). This fixes for any given mass $M_{S}$,
parameter $A$ in  Eq.~(\ref{cross_section}) and average transfer energy
$\langle y\rangle=\langle E_{S}/E_p\rangle$ the normalization of the
initial proton flux, which is marked by $p_{\rm ini}$ in
Fig.~\ref{F4}. The value of the initial proton flux shown in
Fig.~\ref{F4} corresponds for $M_{S}=2$~GeV to $\langle
E_{S}/E_p\rangle\approx 0.1$ and $A\sim 1$.

As background photons in the source we have used infrared/optical photons 
with energies $0.1 {\rm eV} \le \epsilon \le 10$~eV and
number density $n_\gamma = 5 \times 10^{12}/{\rm cm}^3$. This
corresponds to the luminosity $L=5\times10^{43}{\rm erg}/{\rm s}$, if the
radius of the emission region is $R=10^{16}$~cm. After several interactions
with the background photons the accelerated protons lose all their
energy and produce photons and neutrinos. The neutrino flux should obey
the existing experimental limits of AGASA~\cite{agasa_nu},  
Fly's Eye~\cite{Fly'sEye_nu}, RICE~\cite{rice} and GLUE~\cite{glue}, 
which are shown in Fig.~\ref{F4}. 
Photons cascade down to the GeV and sub-GeV region. The existing
diffuse gamma-ray flux measurement by EGRET restricts the photon flux
in the MeV--GeV region, see Fig.~\ref{F4}. However, if high magnetic
fields exists in the source, then part of the photon energy can
cascade down into the sub-MeV region, where the bounds on the diffuse
photon background are at least a factor 10 times weaker than in the GeV
region. Another part of the photon flux can produce large scale
jets~\cite{gamma_jet}, thereby again redistributing energy into the
sub-MeV region.  This uncertainty of the value of the photon flux
makes the existing bounds on the neutrino flux much more important.

Protons escaping from the source at high energies will cascade down to
energies below the GZK cutoff and can contribute to the observed UHECR
spectrum, as shown in Fig.~\ref{F4}. We have supposed that there are no
UHECR sources within $R_{\rm GZK}=100$ Mpc around the Earth. As a result,
the proton spectrum has a sharp cutoff {\it below} $10^{20}$~eV,
see Fig.~\ref{F4}. Thus, if no nearby UHECR sources exist, then even
the HiRes data are inconsistent with the minimal model of protons
coming from uniformly (but rare) distributed UHECR sources. 
 
In Fig.~\ref{F5}, we show the UHE neutrino flux (per flavor) in our
model for two extreme initial proton fluxes: protons with a spectrum
peaked at $E\sim 10^{21}$~eV and for a $1/E^2$ spectrum up to
$E_{\rm max}\sim 10^{21}$~eV. Both cases are consistent with present
experimental limits. In the same figure, we show the sensitivities of
future experiments to neutrino fluxes: the Auger project to electron/muon
and tau-neutrinos~\cite{auger_nu}, the telescope array (TA)~\cite{ta_nu}, 
the fluorescence/\v{C}erenkov detector MOUNT~\cite{mount}, 
and, indicated by squares, the space based OWL~\cite{owl_nu} (we take the
latter as representative also for EUSO), the water-based
Baikal (NT200+)~\cite{baikal_limit},  
ANTARES~\cite{antares} (the NESTOR~\cite{nestor} sensitivity would
be similar to ANTARES according to Ref.~\cite{nu_tele}), the ice-based
AMANDA-II  with sensitivity similar to ANTARES and 
${\rm km}^3$ ICECUBE~\cite{icecube}. We assume that the proposed
water based  ${\rm km}^3$ detectors  
like GVD~\cite{GVD} and NEMO~\cite{NEMO} will have sensitivities 
similar to the one of ICECUBE. 
As one can see in Fig.~\ref{F5}, future
experiments will easily detect UHE neutrino flux in any model of new
light hadrons. In these models, high neutrino fluxes as shown are
unavoidable high---contrary to the case of neutrino produced by UHECR 
protons interacting with CMB photons. In the latter case, the neutrino flux 
can be as high as in Fig.~\ref{F5}, but could be also much 
lower, depending on the initial proton spectrum
and the distribution of sources~\cite{nu_fluxes}.

\section{Discussion: BL Lacs as UHECR sources.}
\label{sec:BLLac}

The results of the previous section do not depend on the particular
type of astrophysical accelerator. However, we have normalized the
astrophysical parameters to those of AGN cores by purpose. First of
all, AGNs are one of the best candidates for proton
acceleration to UHE. Second, a statistically significant correlation of
UHECRs with BL Lacs, a subclass of blazars (AGNs with jets directed to
us) with weak emission lines, was found
\cite{corr_bllac,egret_bllac}. Motivated by this correlation, we discuss  
the particular case of BL Lacertae as sources of new hadrons in this
section.

In Ref.~\cite{corr_bllac}, it was shown that the correlation with BL
Lacs requires a new, neutral component in the UHECR spectrum. Here we
have suggested that this component is due to new, neutral shadrons. 
The trajectories of these shadrons should point towards their
sources, apart from small deflections due to possible magnetic moments.
As we have showed in the previous sections, shadrons 
are a good candidate for UHECRs and can be produced in AGN
cores if the condition~Eq.~(\ref{depth}) is fulfilled.  
Now we address the question if the BL Lacs shown to correlate with
UHECRs in~\cite{corr_bllac,egret_bllac} obey this condition.

As an example, we have checked this condition for BL Lac
RX~J10586+5628, which is 
located at redshift $z=0.144$ and correlated with AGASA doublet $E=
(7.76, 5.35)\times 10^{19}$~eV. First of all, let us note that protons
with these energies can not reach us from the distance $D = 680$ Mpc
(we supposed the actual 'best fit' cosmological model with
$\Omega_M=0.3$, $\Omega_\Lambda=0.7$,  $H_0= 70 {\rm km}/ {\rm s Mpc}$).  
The optical magnitude in the V-band of this object is $15.8$, which
gives as optical luminosity $L = 6\times 10^{44} {\rm erg}/{\rm s}$. 
Since the spectra of BL Lacs are broad in the optical region (see,
e.g., the spectrum of RX~J10586+5628 in~\cite{RXJ10586+5628}), 
the density of 0.5~eV photons is high enough in a region of size
$R=10^{16}$~cm to obey the condition~Eq.~ (\ref{depth}) and thus to
produce shadrons from accelerated protons.

If the UHECR primaries are new particles created in proton
interactions in the source, large secondary neutrinos and photon
fluxes are unavoidable. The neutrino fluxes are too small to be
detectable by current experiments, but photons can cascade down into
the MeV-GeV region in the source, and can be measured. Let us compare
the UHECR flux of the BL Lac RX~J10586+5628 with its gamma-ray 
flux in the MeV-GeV region, measured by EGRET. The two events observed
by AGASA  with energy $E=(7.76, 5.35)\times 10^{19}$~eV allow us to
estimate the integrated UHECR flux, 
$\int dE E F(E) \sim 0.05 {\rm eV/ cm}^2/{\rm s}$, 
while the integrated EGRET flux in the region 100--800~MeV is approximately 
$\int dE E F(E) \sim 10 {\rm eV/ cm}^2/{\rm s}$~\cite{egret_sources}. 
If we suppose that the EGRET flux is mostly due to proton energy
losses, the ratio of fluxes is $5\times 10^{-3}$ -- a value consistent
with our model. However, the comparison above can be considered only as 
an order of magnitude estimate: first, the flux measured by EGRET could
be produced by other interactions. Second, the energy injected by
protons into electromagnetic cascades in the core of  RX~J10586+5628 
can be redistributed out of the line-of-sight and thus not contribute
to the EGRET measurement. The next generation of TeV gamma-ray telescopes, 
like H.E.S.S.~\cite{HESS}, MAGIC~\cite{MAGIC} and VERITAS~\cite{VERITAS}, 
will have sensitivities  in the 10--100~GeV energy region, 
allowing to measure gamma-ray fluxes from distant sources similar to 
BL Lac RX~J10586+5628. Such measurements can be complimentary to the observations
of UHECR from the same objects and will allow to restrict or to
confirm a wide class of UHECR models (including the one we considering
here) which imply the production of secondary particles from protons.

The high optical depth of photons in Eq.~(\ref{depth}) guarantees that
protons lose energy in the interaction region and produce shadrons
with a ratio of $\sigma_{S}/\sigma_{p \gamma}\sim 5\times
10^{-3}$. For an  optical depth of $\tau = 5$, only the fraction
$e^{-5}=6.7 \times 10^{-3}$ of initial protons will escape from the
source without interaction.   
Thus the flux of produced  shadrons is in this case similar to the
flux of escaping UHE protons. This example shows how the same source can
be a source of UHE protons and at same time a source of new UHE
hadrons. If the optical depth $\tau$ is smaller, the source will
dominantly produce protons, if it is higher, it will mostly produce
$S$-hadrons.

It will be interesting to check the UHECR data at lower energy, 
$E\sim (2-4) \times 10^{19}$~eV, for correlations with RX~J10586+5628. 
This would be the typical energy of protons from this object with
$z=0.144$ taking into account energy losses on the way to the 
Earth.  The comparison of the low energy proton flux with a possible
UHE flux of new hadrons can be used to check the consistency of our
model with the assumption that BL Lac are UHECR sources.

In Ref.~\cite{corr_protons}, Tinyakov and Tkachev examined
correlations of BL Lacs with the arrival directions of UHECR allowing for
charges $Q=-1,0,+1$ of the primaries. They showed that the deflection
of charged particles in the galactic magnetic field can significantly
increase the correlation with BL Lacs. If primaries can have
charge $Q=0,+1$, they found that 19 from 57 AGASA events 
correlate with BL Lacs, which have magnitude $m<18$ in optics. 
The probability that this correlation is by chance is $2\times 10^{-4}$.

They assumed that the charged particles are protons and the neutral
ones photons. This interpretation has two important drawbacks:
First, both the highest energy event with $E=2\times 10^{20}$~eV and
charge $Q=+1$, and the event 16 of Table~1 in~\cite{corr_protons} with
energy $E=4.39\times 10^{19}$~eV and charge $Q=+1$, which correlates
with a BL Lac at $z=0.212$, can be only explained as background event.
Second, they were forced to assume that most of the BL Lacs with
unknown redshift are located nearby, $z \le 0.1$.  

Now, if we assume that the UHECR primaries correlated with BL Lacs are
new light hadrons which can have charge $Q=-1,0,+1$, for example
$\tilde B^-$, $\tilde B^0$ and $\tilde B^+$, then the assumptions above are
not required. Shadrons with $Q=+1$ can easily come from high redshift
sources up to $z \sim 0.5$ or even higher. Thus one does not need to
assume that the BL Lacs with unknown redshift are located nearby nor
exclude"unsuitable" sources from the Table~1
in~\cite{corr_protons}. Note also that part of the events with 
$Q=+1$ can still be protons. Also let us remind that the deflection in
the magnetic field in the ultra-relativistic case does not depends on
particle mass and hadrons with $M=2-3$ GeV with charge $Q=+1$ will be
deflected in the same way as protons.

Moreover, the model with light charged hadrons predicts also the
existence of particles with negative charge. However, because these
particles will be produced in the sources due to $p^+ \gamma$
reactions, new particles with $Q=+1$ or $Q=0$ will dominate. Particles
with $Q=-1$ can be produced in such reactions only well above the
production threshold and their expected number should be less than the
number of particles with $Q=+1$ and $Q=0$. Moreover, UHE protons can
increase the number of particles with $Q=+1$ at lower energies. Thus,
a prediction of our model is the existence of a small number of
negatively charged UHE cosmic rays, which average energy is
larger than events with $Q=+1$ and $Q=0$.

It is impossible to check statistically the last statement with
current data, however some hint can be found in
Ref.~\cite{egret_bllac}. The authors of this paper chose as subset of
BL Lacs those which are simultaneously EGRET sources. They found 
that 14 BL Lacs correlate with 65 UHECR  
from AGASA and Yakutsk data, if they allow as charges $Q=+1$ and
$Q=0$. The chance probability of this correlation is $3\times 10^{-7}$
which is more than $5\sigma$ using Gaussian statistics.
In this data set 8 BL Lacs out of 14 are UHECR sources and emit 13 UHECRs.
If one suppose, that UHECR primaries with $Q=-1$ also exist, two more
UHECR will correlate with the {\it same\/} BL Lacs. These two events
have energy $E>5\times 10^{19}$, which is much larger than the average
energy in this data set. Three other UHECR primaries from this data
set can have either $Q=-1$ or $Q=0$. All of 
them also have large energies $E>5\times 10^{19}$. 

One more assumption made by Tinyakov and Tkachev is a cut on the BL Lac
magnitude in the optical range, $m<18$~\cite{corr_bllac,corr_protons}. 
They found that such a cut maximizes the correlation with BL Lacs.
However, they were not able to explain why the correlated BL Lacs are those 
which are most bright in the optical range. In our model of new
particle production, such a criterion is obvious: the optical
background is high enough only in the most brightest BL Lacs. Hence,
only they are able to produce  shadrons in $p\gamma$ reactions. BL
Lacs with lower optical luminosity produce protons, which lose energy
and contribute to the UHECR spectrum at lower energies. Another
interesting hint is the value $m=18$. In Fig.~3 of
Ref.~\cite{nu_sources}, the dependence of the source magnitude as
function of redshift was shown under the condition that sources are
optically thick for protons. This line crosses the value $m=18$ at
redshifts of order $z\sim 0.5-0.6$. This distance is similar to the
one which shadrons with $M=2-3$ GeV can still can propagate. 

Thus, we conclude that the correlation of UHECR with BL Lacertae objects 
which was found in~\cite{corr_bllac} and investigated in detail
in~\cite{corr_protons,egret_bllac} suggests that at least some, if not
most UHECR primaries with $E > 4\times 10^{19}$  
should be {\em new\/} particles with  $Q=-1,0,+1$. 
Explanation of the BL Lacs correlation with $Q=+1$ particles by
protons seems unlikely. The model of new light hadrons, for example 
$\tilde B^-$, $\tilde B^0$ and $\tilde B^+$, naturally explains such
correlation as well as the cut on the BL Lacs magnitude, $m<18$.

\section{Conclusions}
\label{conclusions}

The HiRes experiment published recently their UHECR data which show a
cutoff at the highest energies as expected in the conservative model
of an uniform, continuous distribution of astrophysical sources 
accelerating protons up to energies $E\lsim 10^{21}$~eV. On the other
side, a clustered component in the arrival directions of UHECR with 
$E>4\times 10^{19}$~eV is present in the AGASA data with a statistical
significance of $4.6 \sigma$. If one assumes that this clustered
component is due to point-like astrophysical sources, the predicted
total number of sources of UHECR with $E\ge 4 \times 10^{19}$~eV is of
the order of $400-1200$. In this paper we showed that this number of
sources is so small that the model of continuously distributed protons
sources is a bad approximation at the highest energies. The latter
approximation requires $10^{3-4}$ times more sources than estimated
from the clustering data. In other words, the closest proton source is
located outside the GZK sphere $R>50$ Mpc and the energy spectrum of
UHECR has a sharp exponential cutoff at the energy $E<10^{20}$ eV,
which is inconsistent even with the HiRes data, see Fig.~\ref{F01}.
Including the AGASA data makes this discrepancy even worse.

Moreover, a statistically significant correlation at the level of
$4\sigma$ of the arrival directions of UHECR with BL Lac objects
was found~\cite{corr_bllac}. The closest BL Lacs with known redshift
are located at cosmological distance, $z=0.03$, and protons with $E >
10^{20}$~eV cannot reach us from these sources. Some events at lower
energies also can not be protons, because the redshift of these sources is 
too high. For example, BL Lac RX J10586+5628 is located at $z=0.144$ or at
the distance $700$~Mpc. Protons coming from this object can have
a maximal energy around $2-4\times 10^{19}$~eV, while the 
correlated UHECRs have much higher energies, $E= (7.76, 5.35) \times
10^{19}$~eV.

Our findings above suggest the existence of particles that can be
produced at distant astrophysical objects like BL Lacs, propagate
through the Universe without significant energy losses and produce
air-showers in the Earth atmosphere similar to those of protons.  

In this work we have investigated the possibility that such particles
are new light hadrons. We have showed that such hadrons can be
produced in astrophysical objects in interactions of accelerated
protons with a background of optical photons, if the size of the
interaction region  is larger than the interaction length of protons. 
The interaction of the new hadrons with background photons should  
be suppressed to allow them to escape from the sources without
significant energy losses. This fact as well as the requirement that
the energy losses of the new particles propagating in the CMB are
suppressed compared to protons restrict the new hadrons to be
heavier than 1.5--2~GeV. Since the primary protons also produce
also large neutrino and gamma-ray fluxes, which are bounded by
experimental limits and measurement, only hadrons with masses below
3~GeV are allowed. The possibility to travel over cosmological
distances without decay restrict the lifetime of these particles to be
larger than one month. 
 
As specific example we have considered hadrons containing light bottom
squarks. This case agrees with all existing astrophysical
observations, if the shadron mass is in the window $1.5~{\rm GeV}\lsim
M_S \lsim 3 {\rm GeV}$. Such a new hadron can explain the observation of
UHECR at the highest energies.

If BL Lacs are indeed UHECR sources, our model of new light hadrons
allows to solve several puzzles connected with these
objects. First, all correlated UHECR with zero charge can be our new
hadrons with $Q=0$. Second, our model offers a simple explanation why
only optically bright BL Lacs correlate with UHECRs: only if the
density of optical photons in a BL Lac is high, the probability of
protons to interact and to produce our new hadrons is large enough.
Magnitude $m=18$ can correspond to the redshift $z\sim 0.5-0.6$, a
distance from which our new particles can still reach the Earth
without significant energy losses.

In Ref.~\cite{corr_protons} it was shown that the correlation with BL
Lacs increases if one supposes that some UHECR have non-zero charge.
In particular, a significant correlation was found if some UHECRs have
a positive charge $Q=+1$. It was suggested that these positively
charged particles are protons. However, this assumption forced the
authors of Ref.~\cite{corr_protons} to assume that most of the BL Lacs
with unknown redshift are located at the distances $z<0.1$. 
Furthermore, they had to assume that some of the UHECR which can not
be protons are correlated just by chance. 

These two assumptions can be relaxed in our model if one assumes that
new hadrons with non-zero charge are also long-lived. However, the
existence of such hadrons is disfavored from accelerator experiments.

An important consequence of our model is an unavoidable high UHE
neutrino flux. This flux is well within the sensitivity region of
all future UHECR experiments and can also be detected by ${\rm km}^3$
neutrino telescopes like ICECUBE (or GVD and NEMO). In the case of
an initial proton spectrum $\propto 1/E^\alpha$ with $\alpha\sim 2$ 
even 0.1~km$^3$ neutrino telescopes like AMANDA II, ANTARES
 and NESTOR will be able to detect the diffuse neutrino flux in the
$10^{16}$~eV energy region. 

Another consequence of our model is a cutoff in the UHECR spectrum,
which can be observed around $E\sim 10^{21}$~eV at future UHECR
experiments like the Pierre Auger Observatory, the telescope array and
EUSO. 

New hadrons with 1.5--3~GeV mass can be searched for in existing
accelerator experiments like CLEO and B-factories or with a dedicated
experiment as proposed in~\cite{Berezinsky:2001fy}.

\acknowledgments

We would like to thank Oleg Kalashev for making changes in the
code~\cite{code2} which allowed us to produce the data for 
Figs.~\ref{F4} and \ref{F5}. We are grateful to
Venya Berezinsky, Dmitry Gorbunov, Andrey Neronov, Sergey Ostapchenko,
Sven Heinemeyer,
Georg Raffelt, Peter Tinyakov, Igor Tkachev, and Sergey Troitsky 
for fruitful discussions and comments.
M.~A.~T. thanks the Max-Planck-Institut f\"ur Physik for hospitality and 
the Spanish Ministry of Education and Culture for support through the
grants AP200-1953 and BFM2002-00345.
In Munich, this work was supported by the Deutsche Forschungs\-ge\-meinschaft
(DFG) within the Emmy Noether program and the Sonder\-forschungs\-be\-reich
SFB~375.


\end{document}